\documentclass{appolb}
\usepackage{graphicx}
\usepackage{amsmath}
\usepackage{bm}
\newcommand{\daggerp}{+}

\begin{document}
    \title{Iterative solutions of the ATDHFB equations to determine the nuclear collective inertia
        \thanks{Presented at the 57th Zakopane Conference on Nuclear Physics, {\it Extremes of the Nuclear Landscape}, Zakopane, Poland, 25 August–1 September, 2024.}%
    }

    \author{
        {\underline{Xuwei Sun}}$^{1}$, 
        {Jacek Dobaczewski}$^{1,2}$, 
        {Markus Kortelainen}$^{3}$,
        {David Muir}$^{1}$,
        {Jhilam Sadhukhan}$^{4,5}$,
        {Adrian S\'{a}nchez-Fern\'{a}ndez}$^{1}$,
        and {Herlik Wibowo}$^{1}$ 
        \address{
            $^1$ School of  Physics, Engineering and Technology, University of York,  Heslington, York YO10 5DD, United Kingdom\\[3mm]
            $^2$ Institute of Theoretical Physics, Faculty of Physics, University of Warsaw, ul. Pasteura 5, PL-02-093 Warsaw, Poland\\[3mm]
            $^3$ Department of Physics, University of Jyväskylä, P.O. Box 35, FI-40014 Jyväskylä, Finland\\[3mm]
            $^4$ Physics Group, Variable Energy Cyclotron Centre, 1/AF Bidhan Nagar, Kolkata-700064, India\\[3mm]
            $^5$ Homi Bhabha National Institute, Anushakti Nagar, Mumbai-400094, India}
    }
    \maketitle
    \begin{abstract}
        An iterative adiabatic time-dependent Hartree-Fock-Bogoliubov (ATDHFB) method is developed within the framework of Skyrme density functional theory. The ATDHFB equation is solved iteratively to avoid explicitly calculating the stability matrix. The contribution of the time-odd mean fields to the ATDHF(B) moment of inertia is incorporated self-consistently, and the results are verified by comparing them with the dynamical cranking predictions. The inertia mass tensor is calculated with the density-derivative term evaluated by numerical differentiation.
    \end{abstract}
    
    \section{Introduction}
    Describing nuclear dynamics in terms of collective degrees of freedom enhances our understanding of nuclear rotation, vibration, and fission.
    The nuclear collective inertia measures the resistance of a nucleus to the collective motion [1], which provides valuable insights into nuclear structure.
    
    The physical determination of the inertia against the collective motion requires knowledge of nuclear microscopic dynamics.
    The ideal theoretical framework should describe the collective motion using appropriate collective variables and encapsulate the interactions between individual nucleons within the collective inertia. 
    A widely used microscopic approach to evaluate the moment of inertia is using the Inglis-Belyaev (IB) formula [2, 3].
    However, it has the drawback of neglecting the time-odd mean-field effects, hence underestimating the collective masses by a factor 1.2$\sim$1.4 [4, 5]. 
    In a nucleus, the presence of a local nuclear interaction gives rise to time-odd mean fields through local gauge invariance and spin-spin terms, an effect that should not be overlooked in the analysis of the collective motion.
    
    Assuming that the collective motion is slow compared to the single-particle motion, the fully self-consistent treatment of the time-odd mean field can be obtained with the adiabatic time-dependent Hartree-Fock [6] (ATDHF, without pairing) or Hartree-Fock-Bogoliubov [3, 7] (ATDHFB, with pairing) methods.
    The adiabatic methods yield collective Hamiltonians that depend on quadratic collective velocities and collective inertia, bridging the gap between microscopic many-body theory and phenomenological models based on collective variables. However, explicit solutions are computationally difficult, and more efficient algorithms are very much required.
    In the past, Dobaczewski and Skalski used iterative solutions of the ATDHFB equations to analyze the quadrupole vibrational inertia function in axially deformed samarium and barium nuclei [7].
    Similarly, Li \textit{et al.} developed the ATDHF method based on the expansion of the inertia matrix [8], and Petr{\'\i}k and Kortelainen [9] and Washiyama \textit{et al.} [10] implemented the iterative finite-amplitude method to solve the linear-response equations and determine the time-odd mean fields. 
    
    Recently, we developed a novel method within density functional theory to precisely solve the ATDHF and ATDHFB equations for arbitrarily deformed nuclei. An efficient iterative approach was used to implement the time-odd mean fields exactly, enabling a microscopic and reliable evaluation of inertia for surface vibrations, rotations, and fission.
    
    \section{Formalism}
    The forthcoming paper [5] provides a detailed description of the iterative ATDHFB method. For brevity, here we focus on the formalism of the iterative ATDHF method and neglect pairing.
    
    The time-dependent density can be decomposed as [6]
    \begin{equation}\label{TD_decompose}
        \rho(t) = e^{(i/\hbar)\chi(t)}\rho_0(t)e^{(-i/\hbar)\chi(t)},
    \end{equation}
    where the Hermitian and time-even operators $\rho_0(t)$ and $\chi(t)$ are regarded as \textit{coordinates} and \textit{momenta}, respectively.
    Expanding Eq.~\eqref{TD_decompose} in powers of $i\chi$, the first-order correction to the single-particle density is 
    \begin{equation}\label{drho1_expansion}
        \rho_1 = [i\chi,\rho_0].\\
    \end{equation}
    Applying a similar expansion to the single-particle Hamiltonian $h$ and utilizing the TDHF equation of motion
    $
    i\dot{\rho} = [h,\rho],
    $
    the ATDHF equation can be obtained as
    \begin{equation}\label{ATDHF_EQ}
        i\dot{\rho}_0 = [h_0,\rho_1] + [\Gamma_1,\rho_0],\\
    \end{equation}
    where $h_0$ represents the static single-particle Hamiltonian, and $\Gamma_1$ is the time-odd mean field.
    
    The collective kinetic energy can be calculated from the collective coordinate and momentum, which is
    \begin{equation}
        \mathcal{K} = \frac{1}{2}\textrm{Tr}(\dot{\rho}_0\chi) = -\frac{i}{2}\textrm{Tr}(\dot{\rho}_0[\rho_1,\rho_0]).
    \end{equation}
    As it should depend on the inertia linearly and on the velocity $\dot{q}$ quadratically, that is, $ \mathcal{K} = \frac{1}{2}\mathcal{M}\dot{q}^2$, the collective inertia can be expressed as
    \begin{equation}\label{coll_inerita}
        \mathcal{M}= -\frac{i}{\dot{q}^2}\textrm{Tr}(\dot{\rho}_0[\rho_1,\rho_0]).
    \end{equation}
    
    Here we propose a novel method to determine the time-odd correction to the single-particle density by solving the ATDHF equation iteratively.
    First, we express the ATDHF equation \eqref{ATDHF_EQ} in the Hartree-Fock (HF) single-particle basis of the $h_0$ eigenstates.
    Then the particle-hole (ph) matrix elements of the time-odd density can be derived from the ATDHF equation, which is
    \begin{equation}\label{fixp_R}
        \rho_{1,ph}^{(n+1)} = \frac{1}{\epsilon_p - \epsilon_h}
        \Big[ i\dot{q}\frac{\partial \rho_{0,ph}}{\partial q} - \Gamma_{1,ph}^{(n)} \Big],
    \end{equation}
    where $\epsilon_p$ ($\epsilon_h$) is the particle (hole) energy.
    Since the time-odd mean field $\Gamma_1$ is a functional of the time-odd correction to the density, $\rho_1$, the above equation can be solved by a fixed-point iterative method outlined below.
    
    The procedure begins with a vanishing time-odd mean field
    $\Gamma_1^{(0)} = 0$, whereupon the collective mass equals to the
    IB value. In each iteration, the time-odd density $\rho_1^{(n+1)}$ is
    calculated according to Eq.~\eqref{fixp_R}. Then $\rho_1^{(n+1)}$ is used to
    determine the time-odd mean fields, namely,
    $\rho_1^{(n+1)}\rightarrow\Gamma_1^{(n+1)}$. In each iteration of the latter step,
    $\rho_1$ defines the \textit{adiabatic basis} of its eigenvectors.
    Properties of the adiabatic basis can be established in the following way.
    
    First, we note that
    the definition of $\rho_1$ in Eq.~(\ref{drho1_expansion}) implies that it has only non-vanishing matrix elements ph
    ($\tilde{\rho}_1^{\daggerp}$) and hp ($\tilde{\rho}_1$), that is,
    \begin{equation}\label{rho1a}
        \rho_1= \bigg(
        \begin{array}{cc}
            0 & \tilde{\rho}_1\\
            \tilde{\rho}_1^{\daggerp} & 0
        \end{array}\bigg)
        =
        \bigg(
        \begin{array}{cc}
            0 & U r V^{\daggerp}\\
            V r U^{\daggerp} & 0\\
        \end{array}
        \bigg),
    \end{equation}
    where columns of the $N\times{N}$ and $M\times{N}$ SVD matrices $U$ and $V$, respectively, are normalized and orthogonal ($U^+U=1$ and $V^+V=1$). Here,
    $N$ (the particle number) and $M$ are the dimensions of the HF hole and particle spaces, respectively, and $r$ is a diagonal $N\times{N}$ matrix of positive
    numbers.
    
    Second, we have,
    \begin{equation}\label{rho1b}
        \rho_1\bigg(\!
        \begin{array}{c}
            U\\ \pm V
        \end{array}
        \!\bigg)
        =\bigg(\!
        \begin{array}{cc}
            0 & U r V^{\daggerp}\\
            V r U^{\daggerp} & 0\\
        \end{array}
        \!\bigg)
        \bigg(\!
        \begin{array}{c}
            U\\
            \pm V\\
        \end{array}
        \!
        \bigg)
        = \pm
        \bigg(\!
        \begin{array}{c}
            U\\
            \pm V\\
        \end{array}
        \!
        \bigg) r ,
    \end{equation}
    that is, $2N$ columns
        $\frac{1}{\sqrt{2}}\left(\begin{smallmatrix}
            U\\
            \pm V\\
        \end{smallmatrix}\right)$
        form the adiabatic basis of the $\rho_1$ eigenstates corresponding to eigenvalues $\pm r$.
        The adiabatic basis is thus composed of twice-particle-number ($2N$) ``occupied'' states
        appearing in pairs of opposite ``occupation numbers'' and $M-N$ empty states. This similarity with the standard HF basis
        allows us to easily
        calculate the time-odd spatial densities and currents in close analogy to the standard calculation of the
        time-even densities and fields. Finally, the
        collective inertia $\mathcal{M}^{(n)}$ is evaluated from
        Eq.~\eqref{coll_inerita} at the end of each iteration. The
        fixed-point iteration stops when achieving the desired precision for
        $\mathcal{M}$.
        
        The iterative ATDHF(B) method was implemented in the code HFODD [11, 12], which solves universal non-relativistic nuclear DFT equations in the Cartesian deformed harmonic oscillator basis. The advantage of this efficient method is that it involves only one-body variables and avoids the calculation of the two-body stability matrix [1] entirely, which for deformed superfluid nuclei is a huge and usually numerically prohibitive task. 
        
        \section{Results}
        The iterative ATDHF(B) method is well-suited for determining the collective inertia of arbitrarily deformed superfluid nuclei. In this section, three applications are presented to verify its numerical reliability.
        
        \subsection{Rotational moment of inertia of the axial-deformed nucleus: $^{20}\textrm{Ne}$}
        We validate our method by comparing the ATDHF results with the dynamical cranking (DC) calculations.
        For the instance of rotating along $y$-axis, in the DC calculation, the moment of inertia is obtained by $\mathcal{I}_y = {J_y}/{\omega_y}$, where a small cranking frequency $\omega_y=0.001$ MeV is used to generate the corresponding angular momentum $J_y$ and time-odd densities and mean fields.
        
        \begin{figure}[!htb]
            \begin{minipage}[c]{0.45\textwidth}
                \includegraphics[width=\textwidth]{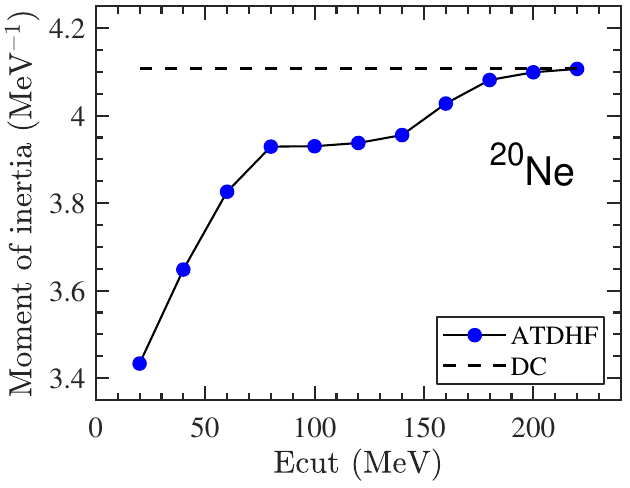}
            \end{minipage}\hfill
            \begin{minipage}[c]{0.5\textwidth}
                \caption{The ATDHF Moment of inertia of $^{20}$Ne (solid line) compared with the value evaluated from the dynamical cranking (DC) calculation (dashed line) for different single-particle cutoffs. The Skyrme interaction SVT is used. \label{Ne20_MOI_Ecut}}
            \end{minipage}
        \end{figure}
        In Fig.~\ref{Ne20_MOI_Ecut}, the calculated ATDHF moment of inertia of $^{20}$Ne is compared with the value obtained from the DC method. 
        The Skyrme interaction SVT [13] is used. 
        As presented in the figure, the ATDHF moment of inertia is very sensitive to the single-particle energy cutoff. 
        To achieve consistency with the DC method, the ATDHF calculation must include all particle states.
        The exact correspondence between the ATDHF and DC moment of inertia in the full single-particle space demonstrates the accuracy of our iterative method.
        
        \begin{figure}[!htb]
            \begin{minipage}[c]{0.45\textwidth}
                \includegraphics[width=\textwidth]{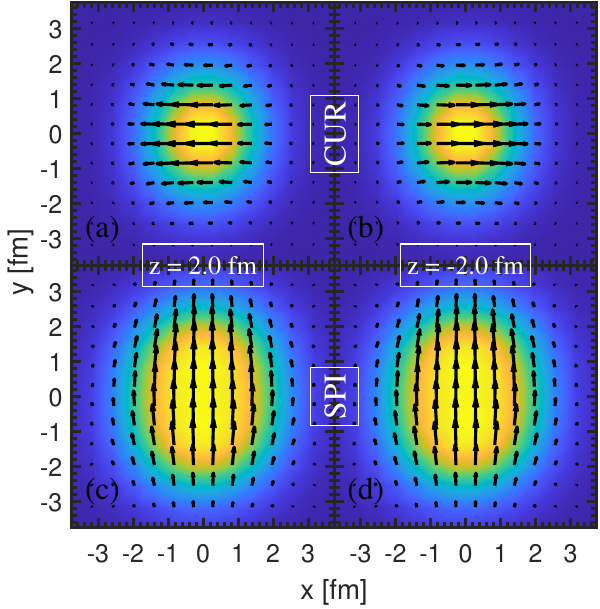}
            \end{minipage}\hfill
            \begin{minipage}[c]{0.5\textwidth}
                \caption{
                    Distributions of the current densities (\protect\ref{current}), panels (a) and (b), and spin densities (\protect\ref{spin}), panels (c) and (d), of $^{20}$Ne rotating along the $y$-axis, projected on the $x-y$ plane at $z=\pm2.0$\,fm. As the nucleus rotates along the $y$-axis, at $z=2.0$ and $z=-2.0$\,fm the current flows in opposite directions.
                    The spin density is mostly aligned along the $y$-axis and being parity-even, at $z=\pm2.0$\,fm it shows symmetric distributions. The colours (in arbitrary units) show the amplitudes of the densities.
                } \label{Ne20_slice}
            \end{minipage}
        \end{figure}    
        The ATDHF method gives a self-consistent treatment to the time-odd mean fields that originate from the time-odd densities and currents. 
        In the coordinate-spin space, the scalar part of the density matrix $\rho(\bm{r}\sigma,\bm{r}'\sigma')$ is defined as
        $
        \rho(\bm{r},\bm{r}')
        =\sum_{\sigma}\rho(\bm{r}\sigma,\bm{r}'\sigma),
        $
        and the pseudo-vector part as
        $
        s(\bm{r},\bm{r}')=
        \sum_{\sigma\sigma'}\rho(\bm{r}\sigma,\bm{r}'\sigma')
        \langle\sigma'|\bm{\sigma}|\sigma\rangle.
        $
        Using these definitions, various time-odd densities and currents can be obtained [14], for example, the vector current density,
        \begin{equation}
            \label{current}
            \bm{j}(\bm{r})=\frac{1}{2i}[(\nabla-\nabla')\rho(\bm{r},\bm{r}')]_{\bm{r}=\bm{r}'},
        \end{equation}
        and pseudo-vector spin density,
        \begin{equation}
            \label{spin}
            \bm{s}(\bm{r}) = \bm{s}(\bm{r},\bm{r}).
        \end{equation}
        
        We analyzed the contribution of each time-odd density or current to the moment of inertia and found the dominant contributions are from the current (\ref{current}) and spin (\ref{spin}) densities.    
        In Fig.~\ref{Ne20_slice}, we show the distributions of current and spin densities of $^{20}$Ne rotating along the $y$-axis.
        
        \subsection{Rotational moment of inertia of the triaxial-deformed nucleus: $^{126}$Ba}
        \begin{figure}[!htb]
            \centering
            \includegraphics[width=0.9\textwidth]{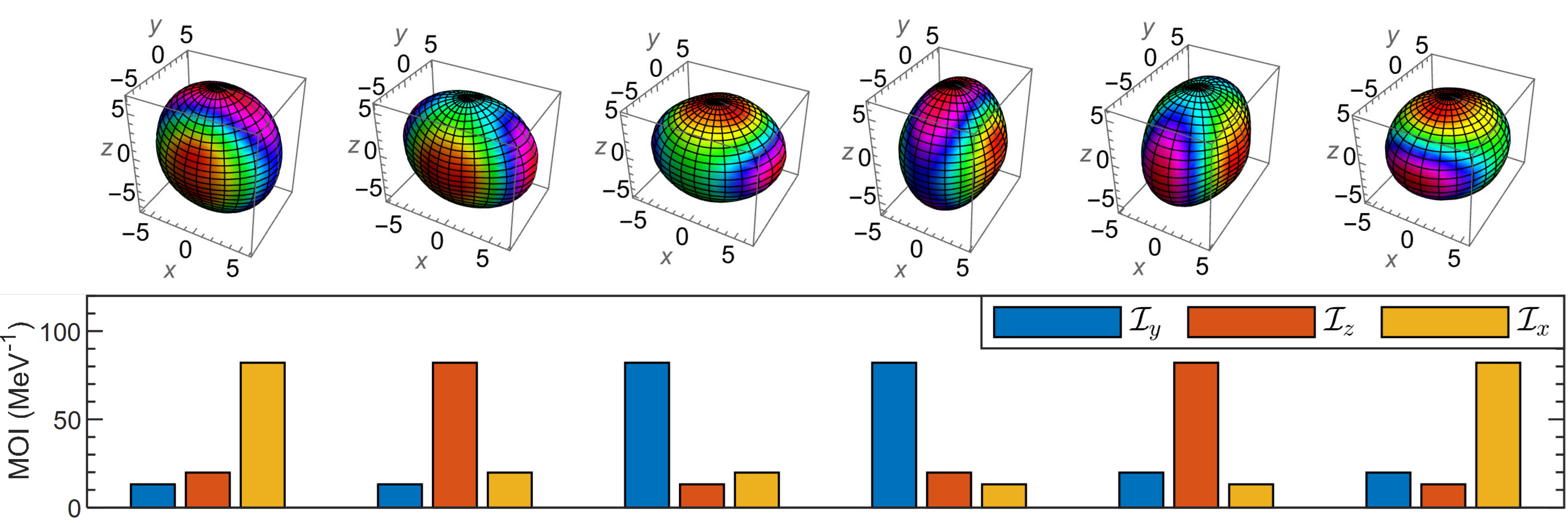}\\	
            \caption{Moments of inertia of $^{126}$Ba rotating along the $y$-, $z$-, and $x$-axes with different orientations. \label{Ba126_MOI}}
        \end{figure}
        The iterative ATDHF method is capable of evaluating the moment of inertia of arbitrarily-deformed nuclei.
        In this section, we investigate the moment of inertia of the triaxial deformed nucleus $^{126}$Ba.    
        The calculations are performed with 16 harmonic oscillator shells. The SVT interaction is used. The calculated intrinsic quadrupole moments of $^{126}$Ba are $Q_{20} =  6.5945$\,b,  $Q_{22} =  5.5582$\,b, which gives $\beta=0.18$ and $\gamma=40.13^{\circ}$.
        Since the nucleus is triaxially deformed, the moments of inertia along the longest, the shortest, and the intermediate axis are different. 
        All the ATHDF moments of inertia are in perfect agreement with the DC results.
        For example, for the rotation along the $y$-axis, $\mathcal{I}_y^{\textrm{ATDHF}} = 13.14124$ and $\mathcal{I}_y^{\textrm{DC}} = 13.1417$\,$\hbar^2$/MeV.
        Both of them are significantly larger than the IB result of 9.84689\,$\hbar^2$/MeV.
        This shows that the time-odd mean fields have a significant impact on the rotational moment of inertia.
        
        When the principal axes of $^{126}$Ba are aligned with the $x$-, $y$-, and $z$-axis, its long, short, and intermediate axes can be oriented in six different spatial configurations.
        In Fig.~\ref{Ba126_MOI}, we show the rotational moments of inertia of $^{126}$Ba in different orientations. 
        The rotational moments of inertia along the nucleus's long, short, and intermediate axis respectively remain constant, regardless of the nucleus's orientation. The invariance validates the self-consistency of the iterative ATDHF method.
        
        \subsection{ATDHFB vibrational inertia of $^{74}$Ge}
        To evaluate the vibrational inertia mass tensor with the iterative ATDHF(B) method, the density-derivative term in Eq.~\eqref{fixp_R} is calculated as
        \begin{equation}
            \frac{\partial \rho}{\partial q}\Big|_{q=q_0} = \lim_{\delta q\to 0}\frac{\rho[q_0 + \delta q]-\rho[q_0]}{\delta q}
        \end{equation}
        with a small difference of the collective variable $\delta q$.
        To examine the accuracy of the above formula, we calculated the quadrupole inertia mass tensor,
        \begin{equation}\label{mass_tensor}
            \mathcal{M} = \Big[
            \begin{array}{ll}
                \mathcal{B}(a_0) & \mathcal{B}(a_0a_2) \\
                \mathcal{B}(a_2a_0) & \mathcal{B}(a_2) \\
            \end{array}\Big],
        \end{equation}
        for $^{74}$Ge, where the collective variable $q$ is $a_0=\langle 2z^2-x^2-y^2\rangle$ and $a_2=\langle x^2-y^2\rangle$.
        
        \begin{figure}[!htb]
            \begin{minipage}[c]{0.5\textwidth}
                \includegraphics[width=\textwidth]{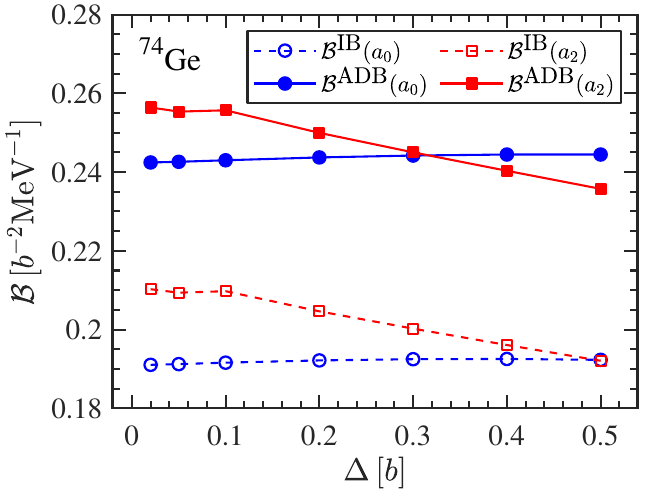}
            \end{minipage}\hfill
            \begin{minipage}[c]{0.45\textwidth}
                \caption{
                    The diagonal components of the IB and ATDHFB (ADB) mass tensor $\mathcal{B}(a_0)$ and $\mathcal{B}(a_2)$ for $^{74}$Ge at $a_0 = 3.5\,b$ and $a_2 = 1.5\,b$. The density-derivative terms are evaluated with positive numerical differences $\delta a_0 = \delta a_2 \equiv \Delta$.  
                    The IB and the ATDHFB inertia tend to stabilize as $\Delta$ decreases from $0.1\,b$ to $0.02\,b$, with uncertainties within 1\%. 
                } \label{Ge74_B20}
            \end{minipage}
        \end{figure}        
        We used the Skyrme interaction SkM* [15] and the volume pairing force $V_t(\bm{r},\bm{r}') =V_0^t \delta(\bm{r}-\bm{r}')\,(t = n,p)$ for $V_0^n=-178.83$ and $V_0^p=-211.20$\,MeV\,fm$^3$.    
        
        Fig.~\ref{Ge74_B20} presents the diagonal components of the IB and ATDHFB mass tensor $\mathcal{B}(a_0)$ and $\mathcal{B}(a_2)$,
        both of them tend to stabilize as $\Delta=\delta a_0 = \delta a_2$ decreases from $0.1\,b$ to $0.02\,b$, with uncertainties within  1\%.
        
        \begin{table}[!htb]
            \centering
            \caption{The off-diagonal component of the IB and ATDHFB mass tensor for $^{74}$Ge. The first and second sign in columns 2 to 5 indicates whether the density-derivative with respect to $a_0$($a_{2}$) is calculated with positive (`+') or negative (`-') difference, respectively.}\label{MT_O}
            
            \begin{tabular}{lcccc}
                \hline
                $\mathcal{B}(a_0a_2)$ [$b^{-2}\text{MeV}^{-1}$] &$-+$ &$++$ &$--$ &$+-$\\
                \hline
                IB &
                0.02842  &0.02714  &0.02960  &0.02835\\
                ATDHFB&
                0.03297  & 0.03144  &0.03447  &0.03296\\
                \hline
            \end{tabular}
        \end{table}
        The off-diagonal component of the mass tensor $\mathcal{B}(a_0a_2)$ depends on the derivatives $\partial\rho/\partial a_0$ and  $\partial\rho/\partial a_2$ simultaneously and is, therefore, more sensitive to $\Delta$.
        Table.~\ref{MT_O} shows the off-diagonal component calculated with $\Delta=0.02\,b$, where both positive and negative values of $\delta a_0$ and $\delta a_2$ with $|\delta a_0| = |\delta a_2| \equiv \Delta$, are used to evaluate the density-derivative terms. The uncertainty caused by the numerical differentiation turns out to be around 4\%. 
        The precision could be further improved by adopting a symmetric, three-point, or five-point differential formula if necessary.
        Since the off-diagonal component is almost one order of magnitude smaller than the diagonal ones, its overall influence on the vibrational inertia is much smaller.    
        
        \section{Conclusions}
        We developed a rapidly converging iterative algorithm to efficiently solve the ATDHF or ATDHFB equations. 
        The method involves only one-body operators, thereby avoiding the calculation of the full stability matrix. 
        The collective rotations in the axial-deformed nucleus $^{20}$Ne and in the triaxial-deformed nucleus $^{126}$Ba were investigated.
        The rotational moment of inertia calculated by ATDHFB method is in perfect agreement with the dynamical cranking method.
        To calculate the inertia mass tensor, the density-derivative term is evaluated using numerical differentiation whereupon 
        the inertia mass tensor is obtained with an accuracy of around 1\%.
        
        
        \noindent
        \\
        \textbf{Acknowledgements}
        \\
        This work was partially supported by the STFC Grant Nos.~ST/P003885/1 and~ST/V001035/1. We acknowledge the CSC-IT Center for Science Ltd., Finland, for the allocation of computational resources. This project was partly undertaken on the Viking Cluster, a high-performance computing facility provided by the University of York. We are grateful for computational support from the University of York High-Performance Computing service, Viking, and the Research Computing team.
        
        \noindent
        \\
        \textbf{References}
        \newline [1] P. Ring and P. Schuck, \textit{The Nuclear Many-Body Problem} (Springer-Verlag, Berlin, 1980).
        \newline [2] D. R. Inglis, Phys. Rev. 103 (1956) 1786.
        \newline [3] S. T. Beliaev, Nuclear Physics 24 (1961) 322.
        \newline [4] J. Libert, M. Girod, and J.-P. Delaroche, Phys. Rev. C 60 (1999) 054301.
        \newline [5] X. Sun \textit{et al.}, to be published.
        \newline [6] M. Baranger and M. Vénéroni, Annals of Physics 114 (1978) 123-200.
        \newline [7] J. Dobaczewski and J. Skalski, Nuclear Physics A369 (1981) 123-140.
        \newline [8] Z. P. Li \textit{et al.}, Phys. Rev. C 86 (2012) 034334.
        \newline [9] K. Petr{\'\i}k and M. Kortelainen, Phys. Rev. C 97 (2018) 034321.
        \newline [10] K. Washiyama \textit{et al.}, Phys. Rev. C 109, L051301 (2014) and references therein. 
        \newline [11] J. Dobaczewski \textit{et al.}, J. Phys. G: Nucl. Part. Phys. 48 (2021) 102001.
        \newline [12] J. Dobaczewski, \textit{et al.}, to be published.
        \newline [13] W. Satuła \textit{et al.}, Phys. Rev. Lett. 106 (2011) 132502.
        \newline [14] Y.M. Engel \textit{et al.}, Nucl. Phys. A249 (1975) 215.
        \newline [15] J. Bartel \textit{et al.}, Nucl. Phys. A386 (1982), 79.
        
    \end{document}